\documentclass[a4paper]{spie}

\usepackage[sc, hang]{caption}
\usepackage{graphicx}
\usepackage{graphics}
\usepackage{makeidx}
\usepackage[pass]{geometry}
\usepackage{booktabs}
\usepackage{marvosym}
\usepackage[hang]{subfigure}
\usepackage{SIunits}
\usepackage{rotating}
\usepackage{csquotes}
\usepackage{url}
\usepackage{multicol}
\usepackage{multirow}
\usepackage{overpic}
\usepackage{amsmath}
\usepackage{commath}
\usepackage{bm}
\usepackage{rotating}
\usepackage{psfrag}
\usepackage{parallel}
\usepackage{fancyhdr}
\usepackage[breaklinks]{hyperref}

\usepackage[english]{babel} 
\usepackage{eso-pic,graphicx}
\usepackage{enumerate}
\usepackage{psfrag}
\usepackage{booktabs}
\usepackage{hyperref}      
\hypersetup{               
   pdftitle={mavisSPIE2022},    
   pdfauthor={GuidoAgapito},  
   colorlinks=true,      
   linkcolor=black,      
   anchorcolor=black,    
   citecolor=black,      
   filecolor=black,      
   menucolor=black,      
   urlcolor=blue,         
   linktocpage=true
}

\usepackage{color, colortbl}
\definecolor{LightGray}{gray}{0.9}
\usepackage{diagbox}
\usepackage{array}
\newcolumntype{C}[1]{>{\centering\let\newline\\\arraybackslash\hspace{0pt}}m{#1}}

\title{MAVIS: performance estimation of the adaptive optics module}

\author[a,c]{Guido Agapito}
\author[b,c]{Daniele Vassallo}
\author[a,c]{Cedric Plantet}
\author[d]{Jesse Cranney}
\author[d]{Hao Zhang}
\author[b,c]{Valentina Viotto}
\author[a,c]{Enrico Pinna}
\author[d]{Francois Rigaut}
\affil[a]{INAF Osservatorio Astrofisico di Arcetri, L. Enrico Fermi 5, 50125 Firenze, Italy}
\affil[b]{INAF Osservatorio Astronomico di Padova, Vicolo dell’Osservatorio 5, 35122, Padova, Italy}
\affil[c]{ADaptive Optics National laboratory in Italy (ADONI)}
\affil[d]{Astralis-AITC - Stromlo, RSAA, Australian National University, Cotter Road, Weston, ACT2600, Australia}

\authorinfo{Further author information:\\Guido Agapito, \Letter
 \hspace{0.5ex} guido.agapito@inaf.it\\}

\begin{document} 

  \maketitle 

\begin{abstract}
The MCAO Assisted Visible Imager and Spectrograph (MAVIS) is a new visible instrument for ESO Very Large Telescope (VLT). Its Adaptive Optics Module (AOM) must provide extreme adaptive optics correction level at low galactic latitude and high sky coverage at the galactic pole on the FoV of 30arcsec of its 4k $\times$ 4k optical imager and on its monolithic Integral Field Unit, thanks to 3 deformable mirrors (DM), 8 Laser Guide Stars (LGS), up to 3 Natural Guide Stars (NGS) and 11 Wave Front Sensors (WFS). A careful performance estimation is required to drive the design of this module and to assess the fulfillment of the system and subsystems requirements. Here we present the work done on this topic during the last year: we updated the system parameters to account for the phase B design and for more realistic conditions, and we produced a set of results from analytical and end-to-end simulations that should give a as complete as possible view on the performance of the system.
\end{abstract}


\section{INTRODUCTION}
\label{sec:intro}

The system design of the Adaptive Optics Module (AOM) of the MCAO Assisted Visible Imager and Spectrograph\cite{2021Msngr.185....7R} (MAVIS) has its roots in a idea from Esposito \textit{et al.}\cite{2016SPIE.9909E..3UE} and it is heavily supported by numerical simulations.
The main parameters of the systems, like the number of Laser Guide Stars (LGS), the number of Deformable Mirrors (DM) and the pitch of the DMs, were selected during the first phase of the design by studies presented in Ref. \citeonline{2020SPIE11447E..1RR,2020SPIE11448E..3RA,2020SPIE11448E..0DV,2020SPIE11448E..6WG,2020SPIE11448E..2LC,2020SPIE11448E..2CZ}.
We proved in these studies that the system is able to fulfill very tight requirements in terms of V band SR -- 10\% on a FoV of 30arcsec diameter (15\% goal) in standard atmospheric conditions at Paranal with bright Natural Guide Stars ($\mathrm{m_J}$=8) -- and sky coverage -- larger than 50\% in the south galactic pole guaranteeing an ensquared energy of at least 15\% on 50mas side in V band (in the same standard atmospheric conditions).
During the second phase of the design the work of performance estimation focused on new simulations with a refined set of the parameters -- we updated some parameters, like the laser spot characteristics and the ones related to the control strategy -- and sensitivity analysis on the various aspect of the atmosphere and of the system.

We relied on a few numerical tools to obtain the results presented in this work: end-to-end simulation tools like PASSATA \cite{doi:10.1117/12.2233963},  COMPASS\cite{2014SPIE.9148E..6OG,2016SPIE.9909E..71G} and YAO\cite{2013aoel.confE..18R}, Fourier-based tools like TIPTOP\cite{2021SPIE11448E..2TN} and other analytical tools like the ones used to estimate sky coverage presented in Ref. \citeonline{2020SPIE11448E..3RA}.
We crosschecked their results to be more confident in their performance prediction: in particular we verified that PASSATA and COMPASS give comparable results in the same configuration within an error of $<<$1\% V band SR.
Then, we refined these tools during this phase to allow us to study new features as we will show in the next sections.

This article is structured in a couple of sections: Sec. \ref{sec:params} focuses on the set of parameters, the control aspects and the baseline performance in the standard atmospheric conditions and Sec. \ref{sec:sens} shows how performance changes as a function of the main parameters and of mis-alignments.

\section{Baseline configuration}\label{sec:params}

In this section we report the parameters used in the baseline configuration. They are summarized in Tab. \ref{tab:params} and they are an update of what we presented in Ref. \citeonline{2020SPIE11448E..3RA}.
In particular we changed the total delay, the NGS pixel scale (see Sec. \ref{sec:LOpixel}) and the parameters of the laser spot.
We have also evaluated the impact of the influence functions of the deformable secondary mirror of the VLT\cite{10.1117/12.2057591} and spider shadows, but we found comparable results with respect to the DM and pupil used in this work. 

We decided to keep two options for the control of the baseline configuration: the first one is the pseudo open loop control with Infinite Impulse Response filters and split tomography as described in Ref. \citeonline{2020SPIE11448E..3RA} and \citeonline{Busoni2019} and the second one is the predictive learn and apply approach reported in Ref. \citeonline{2020SPIE11448E..2LC} and \citeonline{2020SPIE11448E..2CZ}.

One important new feature is the adoption of ad-hoc mis-alignment between SH-WFS sub-aperture grid and the pupil to gain geometrical super-resolution\cite{2022JATIS...8b1514F}: this allows for a reduction of the error of 25nm with respect to the classical approach.

Another new feature is the use of noise priors\cite{Michel-Tallon:2008aa,2010JOSAA..27A...1B,Oberti2019} for the computation of the Minimum Mean Square Error (MMSE) reconstruction matrix.
MAVIS is not affected by significant truncation error, nevertheless it benefits from a more accurate model of the noise covariance matrix: thanks to this approach we are able to reduce the error of 12nm.

We present in Table \ref{tab:budget.HO} the residual WFE breakdown from end-to-end simulations in the baseline configuration for both high and low orders. We quantified each error source individually except for tomographic, generalized fitting and aliasing error that are presented together. The sum of individual HO terms is 107nm, of individual LO terms is 30nm (good NGS asterism with bright stars) and the overall error considering all the error sources (the terms that are not considered in the numerical simulations are reported in Tab. \ref{tab:budget.HO} and in more details in Ref. \citeonline{valentina2022aom}) is 127nm that means a V band SR of 12.3\% ($>$10\% that is the requirement). 

\begin{figure}[h]
    \centering
    \includegraphics[width=0.45\columnwidth]{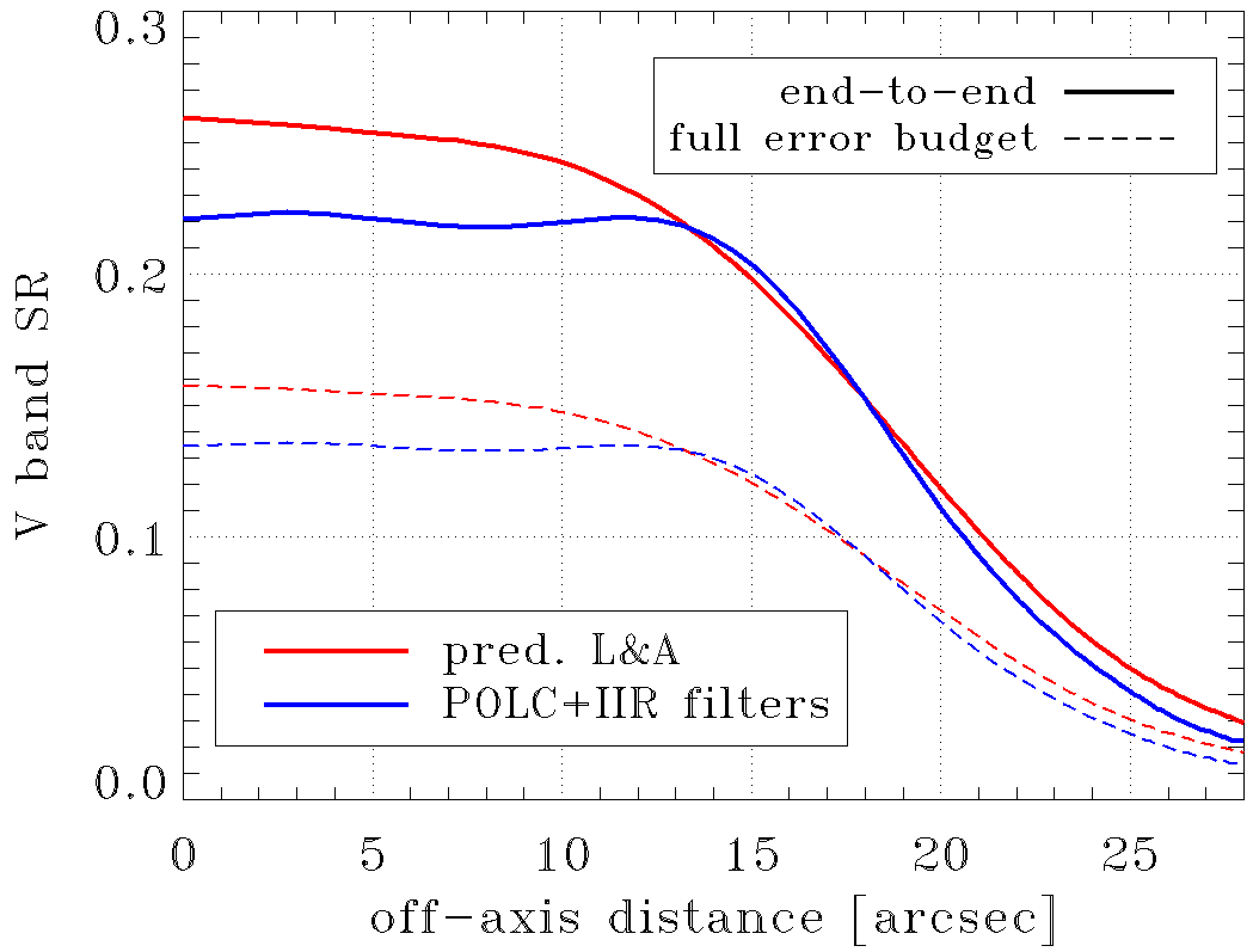}
    \caption{V band SR as a function of off-axis angle for the two control options of MAVIS (see Tab. \ref{tab:params}) considering the output of the end-to-end simulations and the full error budget.}
    \label{fig:MAVIS_SRV}
\end{figure}

Please note that when the predictive learn $\&$ apply control is used the performance improves because the temporal error (38nm, see Tab. \ref{tab:budget.HO}, that means a ratio of 0.83 of V band SR)
(see Tab. \ref{tab:budget.HO}) is reduced as can be seen in Fig. \ref{fig:MAVIS_SRV}.
Please note that in this figure the V band SR with the full error budget (blue dashed line) is slightly higher than the one presented above in the text (about 1\%).
We think that the discrepancy is mainly due to the finite simulated time, 5s, and to the finite sampling of the pupil (0.0222m/pix).
In particular this second features produces a slightly smaller fitting error of 10nm.
Another 10nm are due to the approximation of the DM pitch: in the error budget we considered a pitch of 0.22m, while in the simulation the exact value is 0.216m.
This shows how much the V band SR is sensible to small errors and approximations that have typically no impact on systems with a larger error budget.

\begin{figure}[h]
    \centering
    \subfigure[V band ensquared energy in 50mas diameter.\label{fig:sky_cov_EE}]
    {\includegraphics[width=0.48\columnwidth]{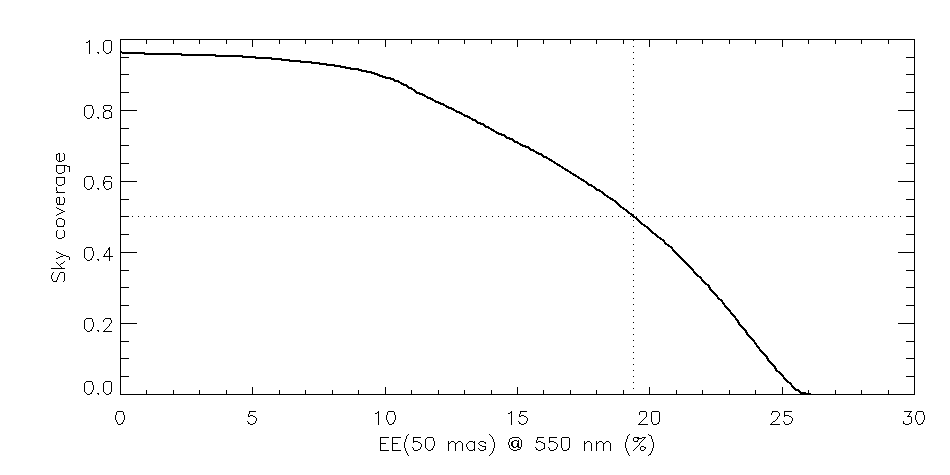}}
    \subfigure[V band FWHM.\label{fig:sky_cov_FWHM}]
    {\includegraphics[width=0.48\columnwidth]{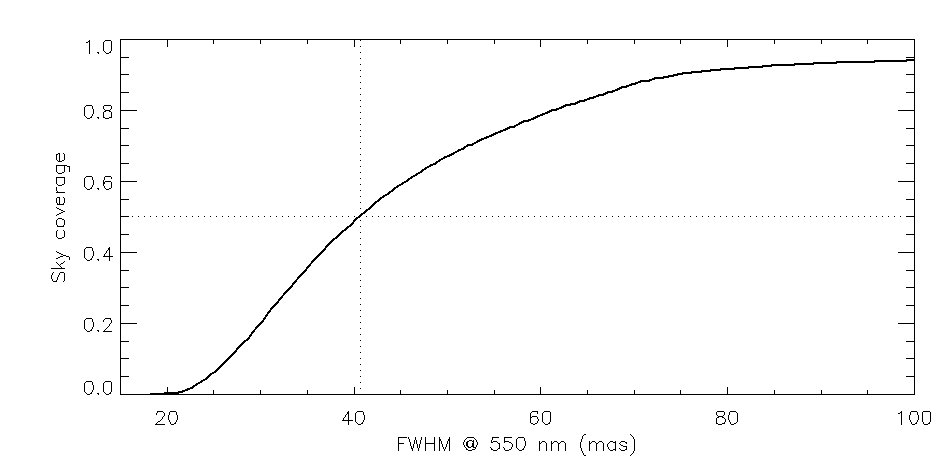}}
    \caption{Sky coverage at the South Galactic Pole.\label{fig:sky_cov}}
\end{figure}

Then in terms of sky coverage we get an Ensquared Energy (EE) in 50 mas at 550 nm of more than 19.4\% for half of the pointings at the South Galactic Pole (Fig. \ref{fig:sky_cov_EE}), which is above the requirement of EE = 15\% for half of the pointings. This corresponds to a Full Width at Half Maximum (FWHM) less or equal to 40.7 mas (Fig. \ref{fig:sky_cov_FWHM}). This result takes also into account the extra error of 58 nm (see Tab. \ref{tab:budget.HO}) as a SR reduction on the NGSs, following the Mar{\'e}chal approximation. The method to compute sky coverage is described in Agapito \textit{et al.} 2020 \cite{2020SPIE11448E..3RA}. Note that the computation of low-order residuals for a given NGS asterism is the same as in the TIPTOP software \cite{2021SPIE11448E..2TN}.

%
%
%
\begin{table}[h]
\caption{Summary of the baseline MAVIS parameters (in simulation).}
\label{tab:params}
\begin{center}
\begin{small}
	\begin{tabular}{|l|c|}
		\hline
		\textbf{Parameter} & \textbf{value}\\
		\hline
		Telescope diameter &  8m (sampled by 360pixels)\\
		Central obstruction & 1.25m (DSM wind screen external diameter)\\
		Pupil mask & round, no spiders\\
		Zenith angle & 30deg\\
		Science FoV (diameter) & 30arcsec\\
		Technical FoV (diameter) & 120arcsec\\
		\hline
		Atmospheric turbulence & 1 profile with 10 layers (Paranal median)\\
		$r_0$ & 0.126m\\
		$L_0$ & 25m\\
	    \hline
	    NGS full throughput & 0.22\\
		LGS full throughput (with laser splitting) & 0.18\\
	    \hline
		Ground DM & conjugated at 0m with $\sim$0.22m pitch (circular act. grid)\\
		Post focal DM no 1 & conjugated at 6000m with $\sim$0.25m pitch\\
		Post focal DM no 2 & conjugated at 13500m with $\sim$0.32m pitch\\ 
		\hline
		NGS WFS number & 3\\
		NGS WFS off-axis angle (good asterism) & 20arcsec\\
		NGS WFS nSA & 1 or 2\\
		NGS WFS pixel scale & 20 (1SA) or 40 (2SA) mas\\
		NGS WFS detector RON & 0.5 $\mathrm{e^-/pixel/frame}$\\
		NGS WFS detector dark current & 20 $\mathrm{e^-/pixel/s}$\\
		NGS (H band) sky background & 2100 $\mathrm{e^-/m^2/arcsec^2/s}$ (no moon)\\
		NGS flux (full aperture, good asterism) & 1000 ph/ms\\
		\hline
		LGS WFS number & 8\\
		LGS WFS off-axis angle & 17.5 arcsec\\
		LGS WFS nSA & 40$\times$40\\
		LGS WFS FoV & 5.0 arcsec\\
		LGS WFS pixel scale & 0.866 mas\\
		LGS WFS detector RON & 0.2 $\mathrm{e^-/pixel/frame}$\\
		LGS WFS detector charge diffusion FWHM & 0.5pixel \\
		LGS launcher number & 4\\
		LGS launcher off-axis distance & 5.5m\\
		LGS flux per sub-aperture (0.04 m$^2$) & 75ph/ms\\
		LGS rotations (for super res.)&  [133.7, 118.7, 61.3, 46.3, 313.7, 298.7, 241.3, 226.3] deg \\
		LGS shifts (for super res.)&  RMS = 0.1SA and absolute values $\leq$ 0.2SA\\
		\hline
		Sodium profile & ``multi peak''\cite{2014A&A...565A.102P} with short axis of 1.35arcsec\cite{haguenauer2022}\\
		Laser jitter & 150mas RMS \\
		\hline
		Control (option 1) & POLC with split tomography\cite{Busoni2019}, IIR filters\cite{2020SPIE11448E..3RA} and noise priors\cite{Michel-Tallon:2008aa,2010JOSAA..27A...1B,Oberti2019}\\
		Control (option 2) & predictive learn \& apply\cite{2020SPIE11448E..2LC,2020SPIE11448E..2CZ}\\
		Centroiding algorithm & CoG (LGS), windowed CoG (NGS)\\
		Framerate & 1000Hz\\
		Total delay & 2.6ms\\
		\hline
		\multicolumn{2}{c}{\scriptsize Note: DSM is Deformable Secondary Mirror, NGS is Natural Guide Star, LGS is Laser Guide Star, DM is Deformable Mirror,}\\
		\multicolumn{2}{c}{\scriptsize WFS is Wavefront Sensor, SA is Sub-Aperture, FoV is Field of View, RON is Read-Out Noise, POLC is pseudo-open loop }\\\multicolumn{2}{c}{\scriptsize control, IIR is Infinite Impulse Response and CoG is Center of Gravity.}\\
	\end{tabular}
\end{small}
\end{center}
\end{table}
\begin{table}[h]
  \caption{Breakdown of MAVIS AO residual wavefront error.}
  \label{tab:budget.HO} 
  \smallskip
  \begin{minipage}{.5\linewidth}
    \centering
    \begin{tabular}{| l | c |}
      \hline
      \multicolumn{2}{| l |}{\textbf{High Orders}}\\\hline
      Error term & Error [nm]\\\hline
      High-frequency fitting\cite{Rigaut_1998} & 65.3\\ \hline
      Tomogr. + gen. fitt. + alias. & 58.1\\ \hline
      Measurement noise & 40.5\\ \hline
      Temporal & 37.9\\ \hline
      Sodium elongation/truncation & 26.4\\ \hline
      LGS jitter & 5.4\\ \hline
    \end{tabular}\smallskip
  \end{minipage}
  \begin{minipage}{.5\linewidth}
    \centering
    \begin{tabular}{| l | c |}
      \hline
      \multicolumn{2}{| l |}{\textbf{Low Orders (bright)}}\\\hline
      Error term & Error [nm]\\\hline
      Tomographic & 27.9\\ \hline
      Measurement noise & $\sim$0\\ \hline
      Temporal & 11.3\\ \hline
    \end{tabular}
    \begin{tabular}{| l | c |}
      \hline
      \multicolumn{2}{| l |}{\textbf{Other errors}}\\\hline
      Error term & Error [nm]\\\hline
      Extra & 58.0\\ \hline
      Vibrations (all modes) & 21.0\\ \hline
    \end{tabular}\smallskip
  \end{minipage}
\end{table}

\section{SENSITIVITY ANALYSIS}\label{sec:sens}

\subsection{Sensitivity to atmospheric profile}\label{sec:srMap}

\begin{figure}[h]
    \centering
    \subfigure[Average V band SR map.\label{fig:srMap1}]
    {\includegraphics[width=0.48\columnwidth]{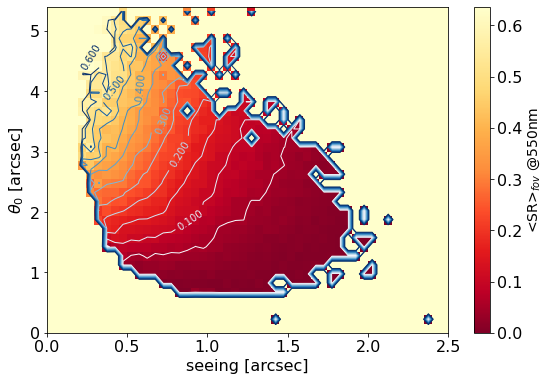}}
    \subfigure[Standard deviation of the V band SR map.\label{fig:srMap2}]
    {\includegraphics[width=0.49\columnwidth]{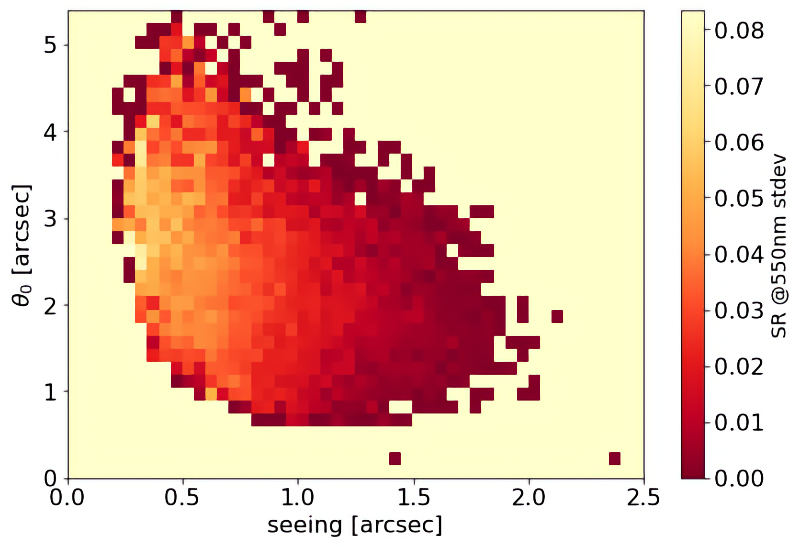}}
    \caption{V band SR (average on science FoV) map as a function of seeing and $\theta_0$. Bins size is 0.05arcsec seeing $\times$ 0.15arcsec  $\theta_0$. The values reported are the average SR for all the profiles that fall in the bin.\label{fig:srMapAll}}
\end{figure}

In this section we present how MAVIS performance changes with the atmospheric profile.
We based this analysis on the release 2019B of the Stereo-SCIDAR profiles\cite{2020SPIE11448E..1WB,10.1093/mnras/sty1070}: we compute the average V band SR in the science FoV for 19356 profiles with TIPTOP and we summarize the results in Fig. \ref{fig:srMapAll}.
Maximum SR values can be found in the top-left part of the plot, where seeing is small and $\theta_0$ is large, and minimum SR values can be found in the bottom-right part, where seeing, $\epsilon$, is large and $\theta_0$ is small.
Iso-performance curves are oriented approximately as lines $\theta_0-k \epsilon = 0$.

We did a further analysis considering these profiles: we selected 529 profiles with seeing close (±1\%) to the median value, 0.72arcsec, and we computed tomography and general fitting error in the science FoV, $\Phi$=30arcsec, for combination of post focal altitudes 4-8 and 12-18km.
We found that best conjugation altitudes are 5 and 14km, close to the current baseline values, and sensitivity is small with a large combination of altitudes close to an average error of 60nm.

\subsection{Sensitivity to zenith angle and seeing}

\begin{figure}[h]
    \centering
    \subfigure[Sensitivity to zenith angle. Error variance changes following approximately a power of 5/3 of the airmass.\label{fig:sens_zenith}]
    {\includegraphics[width=0.43\columnwidth]{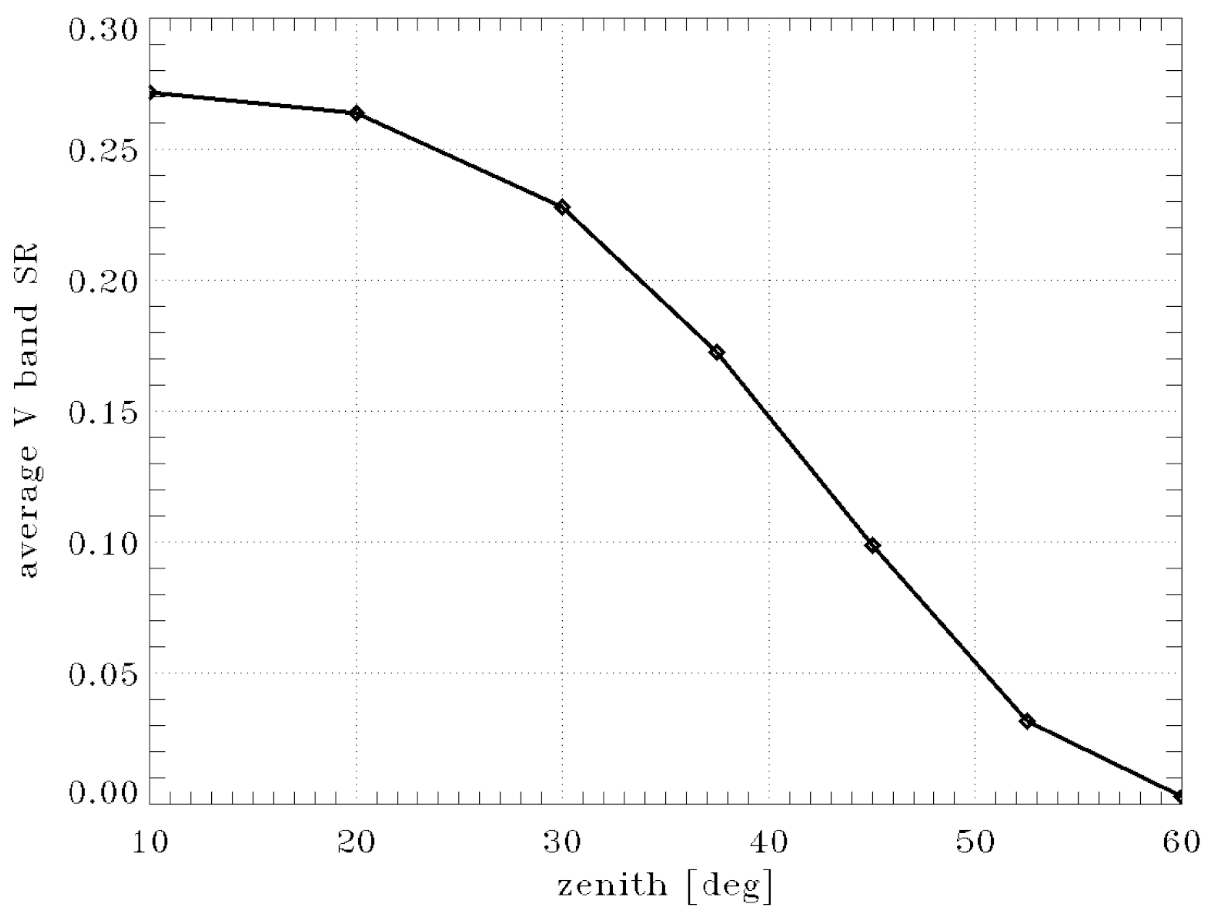}}
    \subfigure[Sensitivity to seeing value (each point is the average of 5 atmospheric and noise realizations). Error variance changes following approximately a power of 5/3 of the seeing value.\label{fig:sens_seeing}]
    {\includegraphics[width=0.42\columnwidth]{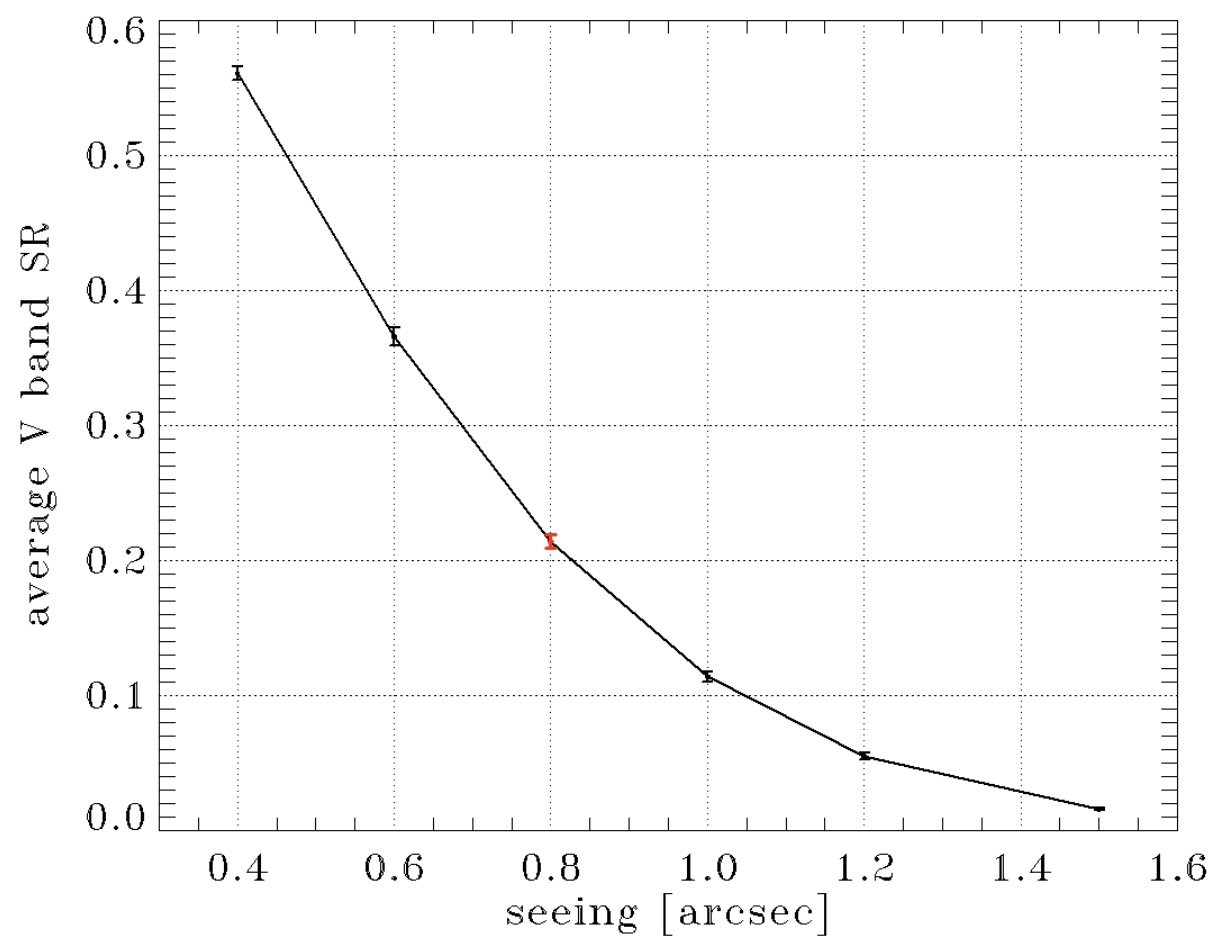}}
    \caption{V band SR as a function of zenith angle and seeing.\label{fig:sens_zen_see}}
\end{figure}

Sensitivity to zenith angle and seeing was evaluated with end-to-end simulations. We found that MAVIS is strongly sensitive to these parameters:
\begin{itemize}
    \item Going from 50 to 10deg of zenith we gain a factor 9 in V band SR (see Fig. \ref{fig:sens_seeing}). Performance changes because effective seeing is a function of the airmass, but also because distance from the pupil of the atmospheric layers is a function of the airmass. In fact, atmospheric layers conjugation altitude impacts generalized fitting and tomographic error. Note that error variance changes following approximately a power of 5/3 of the airmass.
    \item From the worst seeing case to the baseline one there is a factor 20 on V band SR (see Fig. \ref{fig:sens_zenith}). Performance change is dominated by the fitting error and error variance changes following approximately a power of 5/3 of the seeing value.
\end{itemize}
We think that these results are a good occasion to emphasize once again the fact that V band SR is highly sensitive to the simulation parameters.

\subsection{Sensitivity to $\theta_0$}

\begin{figure}[h]
    \centering
    \includegraphics[width=0.90\linewidth]{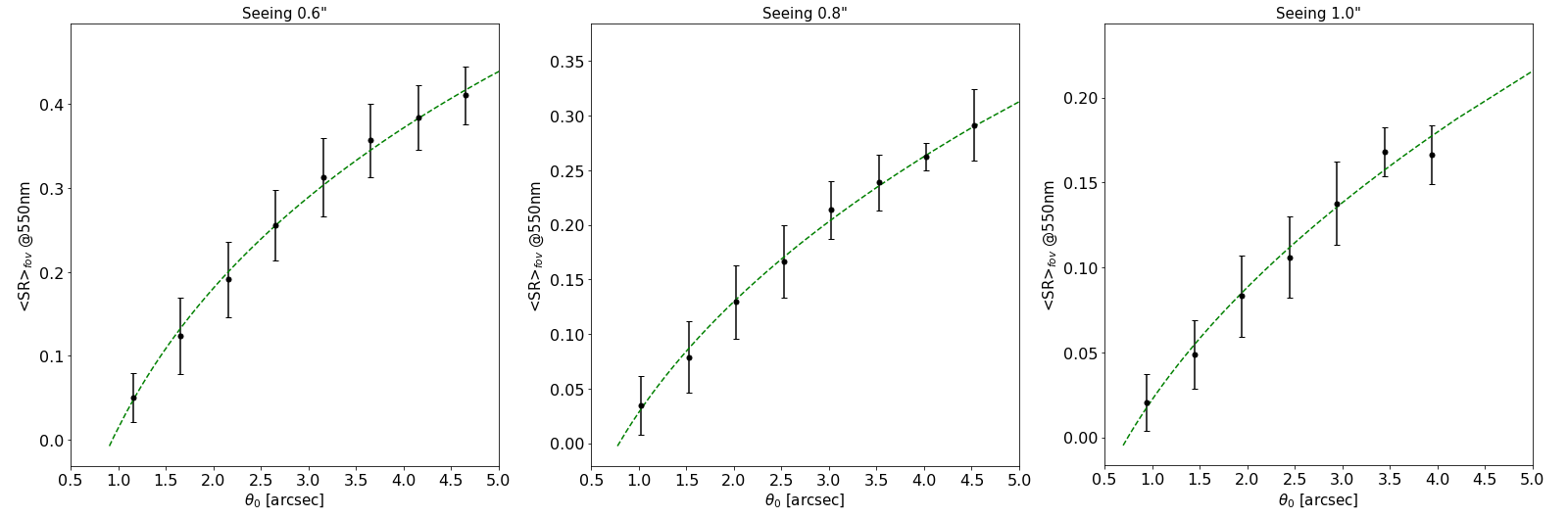}
    \caption{V band SR as a function of $\theta_0$ from the analysis presented in Sec.\ref{sec:srMap} (each point is the average of the atmospheric profiles with $\theta_0$ in bins of 0.5arcsec) for three different seeings: 0.6, 0.8 and 1.0arcsec. Error variance follows approximately a power of $-4/5$ of the $\theta_0$ value.}
    \label{fig:SR_vs_theta0_bins_05}
\end{figure}

We derived sensitivity to $\theta_0$ from the analysis presented in Sec.\ref{sec:srMap}.
We chose those profiles having seeing in the interval $\pm$2\% of a given input seeing: 0.6, 0.8 and 1.0arcsec.
We binned the error as a function of isoplanatic angle at steps of 0.5arcsec and fit a power law to the data (see Fig.~\ref{fig:SR_vs_theta0_bins_05}): we found that the error variance changes following approximately a power of -4/5 of the $\theta_0$ value.
%

\subsection{Sensitivity to wind speed}

Sensitivity to this parameter is relatively low with respect to other atmospheric parameters, nevertheless without a predictive control the V band SR could decrease by a factor 2 going from the best conditions to the bad ones.
Error follows approximately a power of 2 of the average wind speed because no specific optimization is used here and most of the atmospheric turbulence is below the bandwidth of the temporal filters.
\begin{figure}[h]
    \centering
    \includegraphics[width=0.45\linewidth]{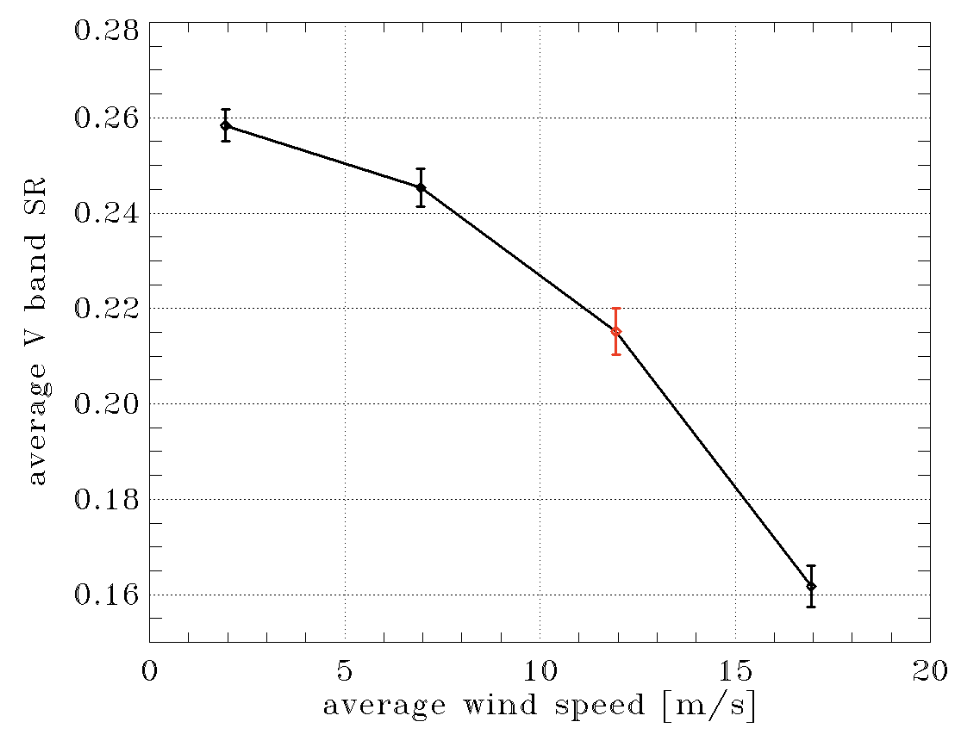}
    \caption{V band SR as a function of wind speed (each point is the average of 5 atmospheric and noise realizations). Error variance follows approximately a power of 2 of the average wind speed.}
    \label{fig:sens_wind}
\end{figure}

\subsection{LGS return flux}

We consider here the performance sensitivity to LGS flux. Sensitivity to this parameter is relatively low, in particular when the most probable values are considered: baseline value is 3500ph/ms/m$^2$, while 90 percentile value is approximately half of this value\cite{haguenauer2022}. We have a reasonable margin in case of the requirement of 10\% V band SR (see dashed line at 1750ph/ms/m$^2$ it is about 12\%), and with good return flux it is possible to fulfill the goal of 15\% V band SR.
Performance does not scale simply as a function of the square root of the flux because noise covariance matrix and temporal filter gains are optimized for each flux considered.
\begin{figure}[htbp]
    \centering
    \includegraphics[width=0.45\linewidth]{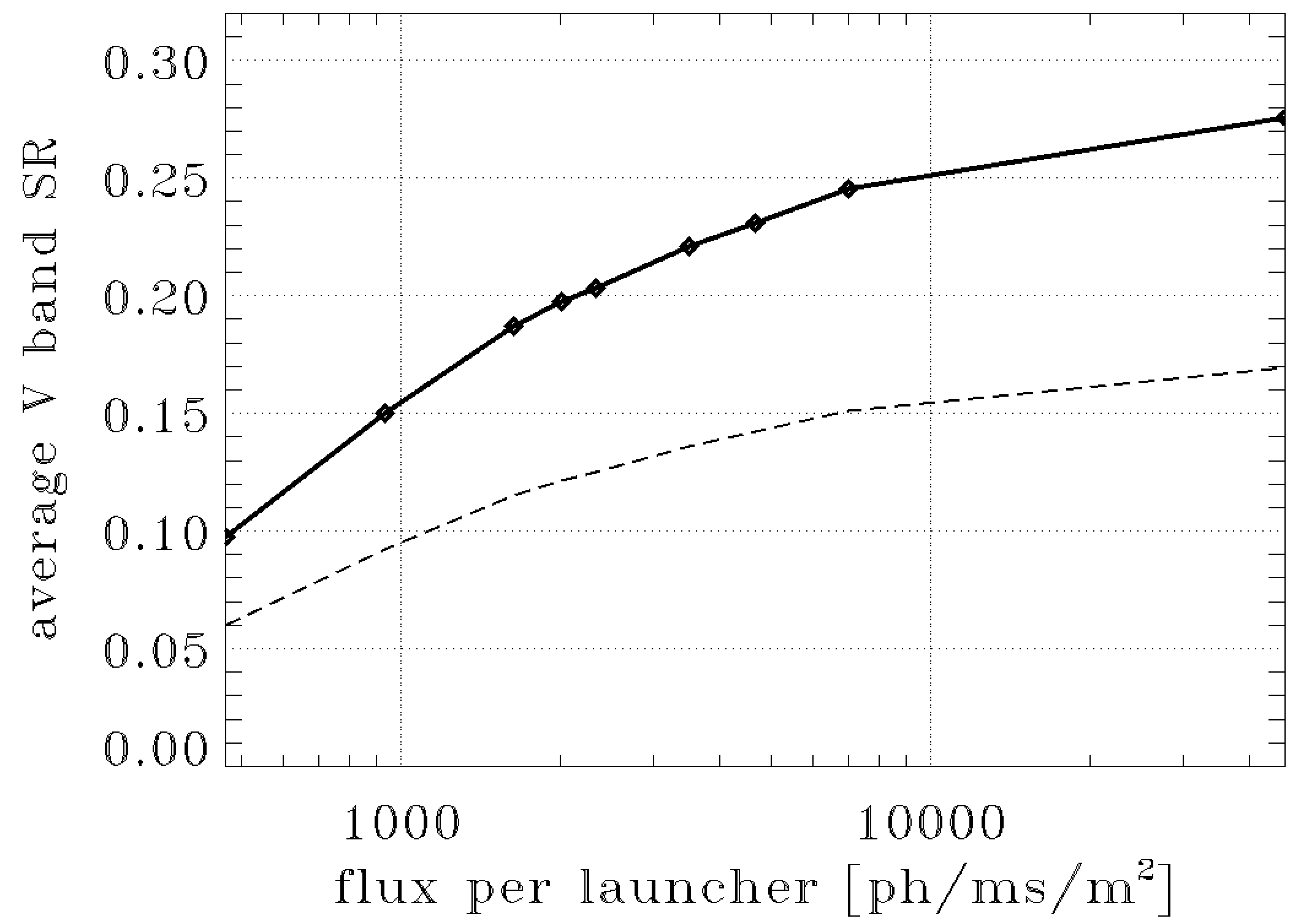}
    \caption{V band SR as a function of laser return flux, dashed line is considering the full error budget. Please note: from ph/ms/m$^2$ to ph/s/m$^2$ at M1 entrance: multiply by 2777, given by reciprocal of full throughput (1/0.18), the splitting factor (0.5) and conversion in seconds (1000).}
    \label{fig:Flux_performance}
\end{figure}

\subsection{LGS spot size and jitter}
 
We studied the sensitivity to the main parameters of the LGS spot: size of the major axis (sodium elongation), size of the minor axis (uplink propagation) and jitter (residual tip-tilt).
We already know\cite{2020SPIE11448E..3RA} that MAVIS has a relatively low sensitivity to sodium elongation when the most elongated spots fall on the diagonal of the sub-aperture, with a difference of a few nm of errors between ``multi-peak'' and ``very wide'' sodium profiles\cite{2014A&A...565A.102P}.
Then, in the last months, we introduced noise priors\cite{Michel-Tallon:2008aa,2010JOSAA..27A...1B,Oberti2019} that are able to further reduce this sensitivity and we focused on the value of the LGS WFS FoV: we verified that there is no improvement in increasing it from 5.0 to 5.5arcsec.

\begin{figure}[h]
    \centering
    \subfigure[V band SR as a function of sodium laser spot minor axis.\label{fig:short_axis}]
    {\includegraphics[width=0.44\columnwidth]{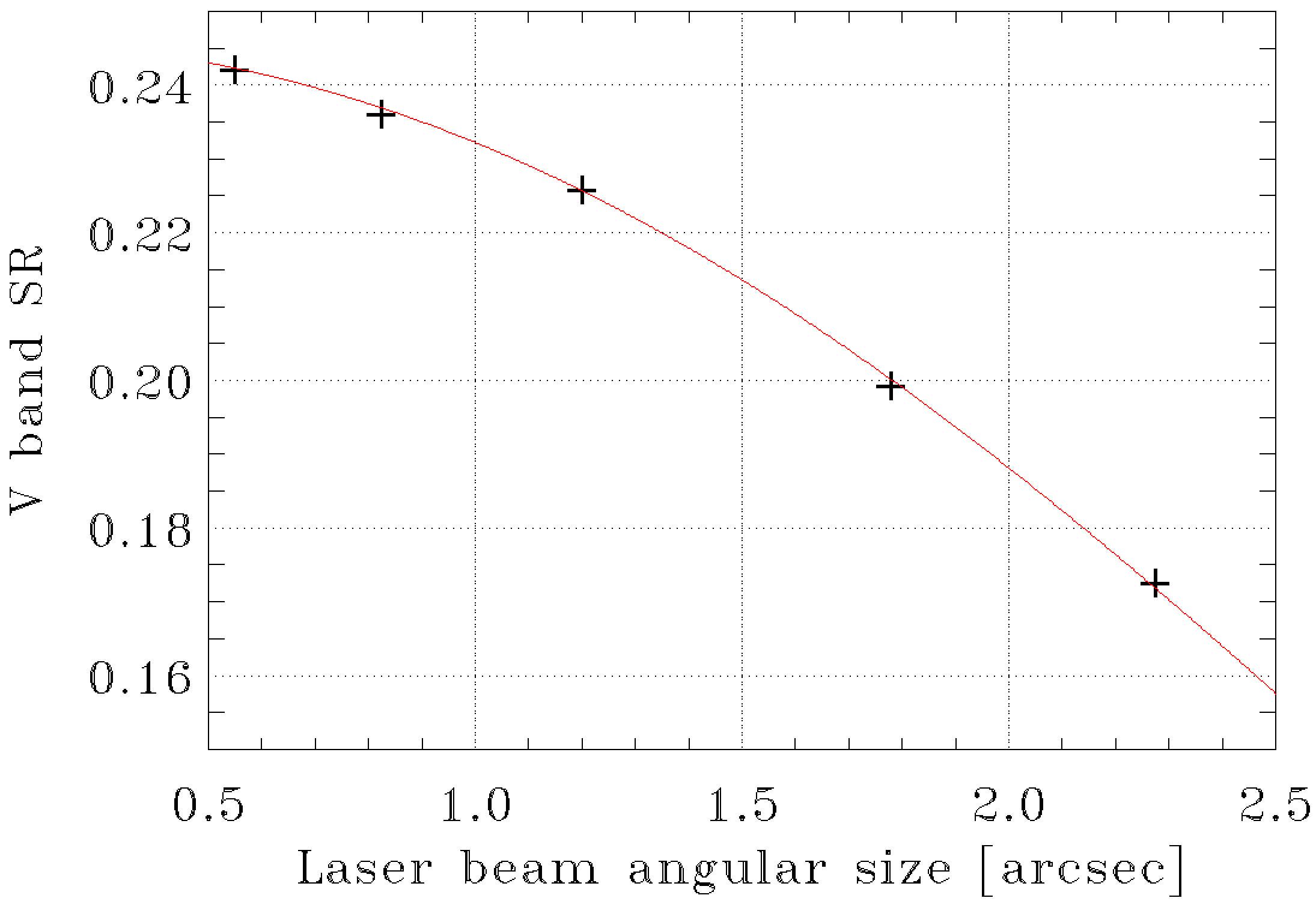}}
    \subfigure[V band SR as a function of sodium laser jitter and for different spot minor axis values ($\phi$).\label{fig:jitter}]
    {\includegraphics[width=0.42\columnwidth]{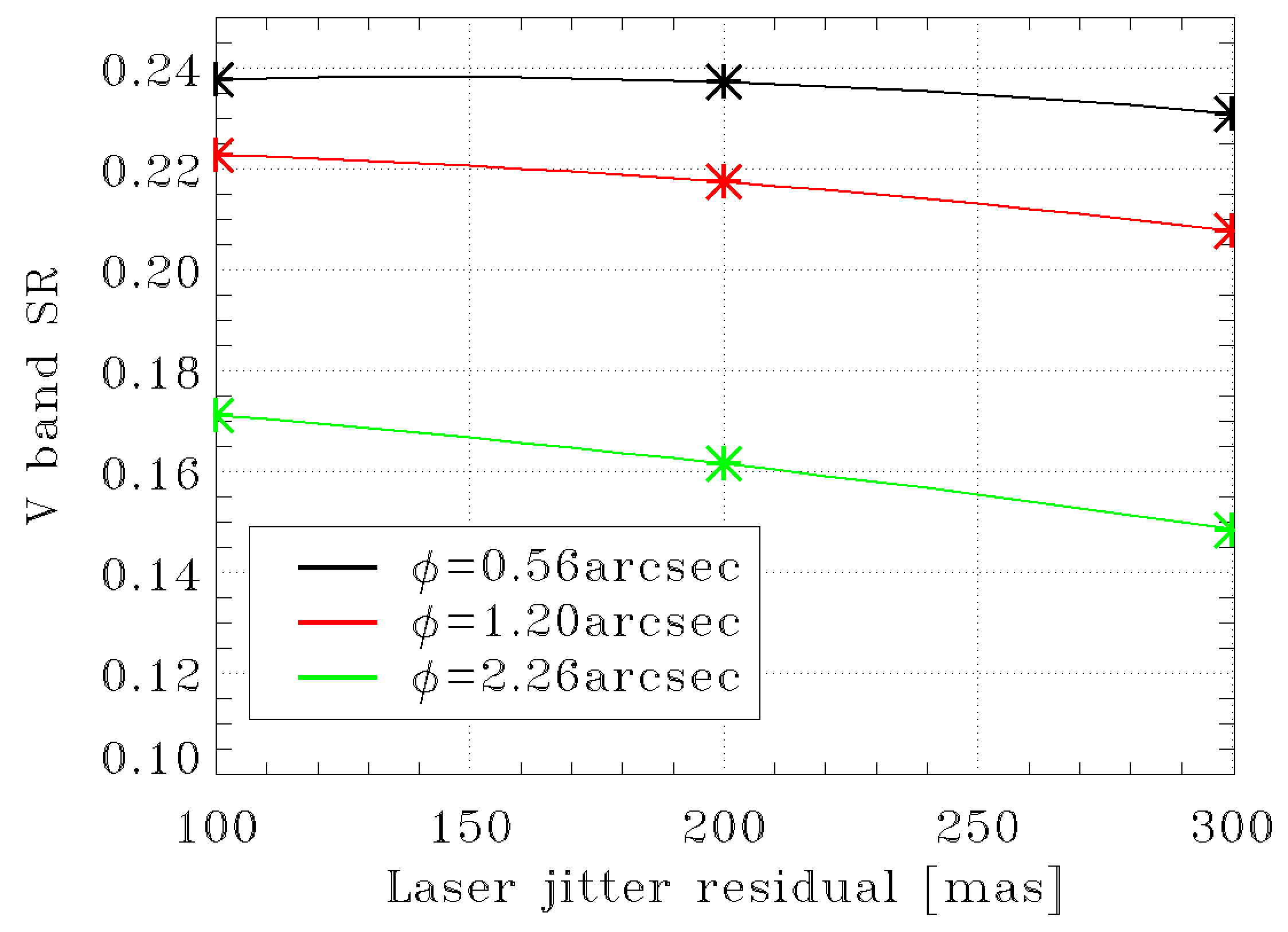}}
    \caption{Sensitivity to laser parameters.\label{fig:laser}}
\end{figure}

Then we focused on sodium laser spot minor axis and laser jitter.
Sensitivity to laser spot minor axis is reported in Fig.~\ref{fig:short_axis}. Here we expect values of about 1.35arcsec\cite{haguenauer2022}.
Then, we simulated uplink propagation and jitter compensation using the tip-tilt mirror coupled to each launcher, with bandwidths from 100 to 1000Hz and seeing values up to 1.5arcsec, and considering that a single mirror has to correct for both LGS generated by a single launcher. We found that best case residual jitter is of the order of 100mas and worst case is of the order of 300mas (minimum and maximum abscissa values of Fig.~\ref{fig:jitter}).
Note that the sensitivity to laser jitter is used to define requirements to the bandwidth of the jitter compensation loop and to the jitter mirrors. 
%
%
%
%

\subsection{LO WFS pixel scale}\label{sec:LOpixel}

\begin{figure}[h]
    \centering
    \subfigure[Curve considering the 2$\times$2 configuration, 500Hz framerate and SR(H)=0.8 on the NGS star.\label{fig:noise_LO_2x2}]
    {\includegraphics[width=0.48\columnwidth]{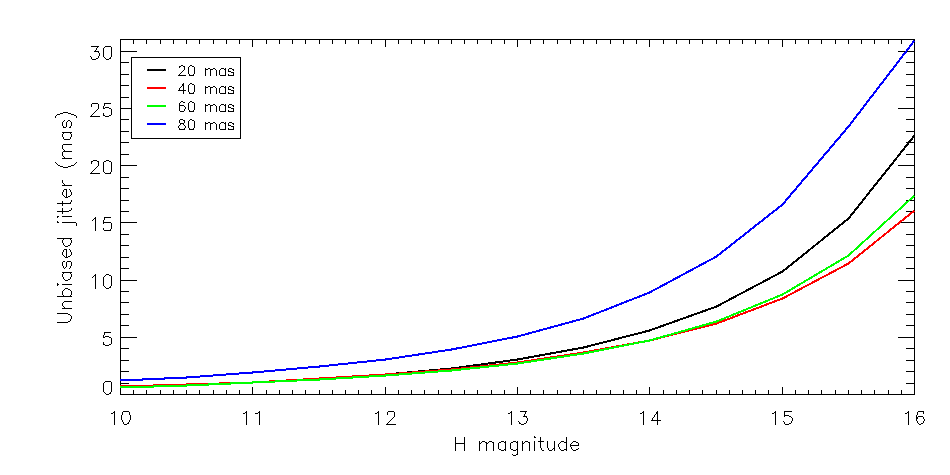}}
    \subfigure[Curve considering the 1$\times$1 configuration, 100Hz framerate and SR(H)=0.2 on the NGS star.\label{fig:noise_LO_1x1}]
    {\includegraphics[width=0.48\columnwidth]{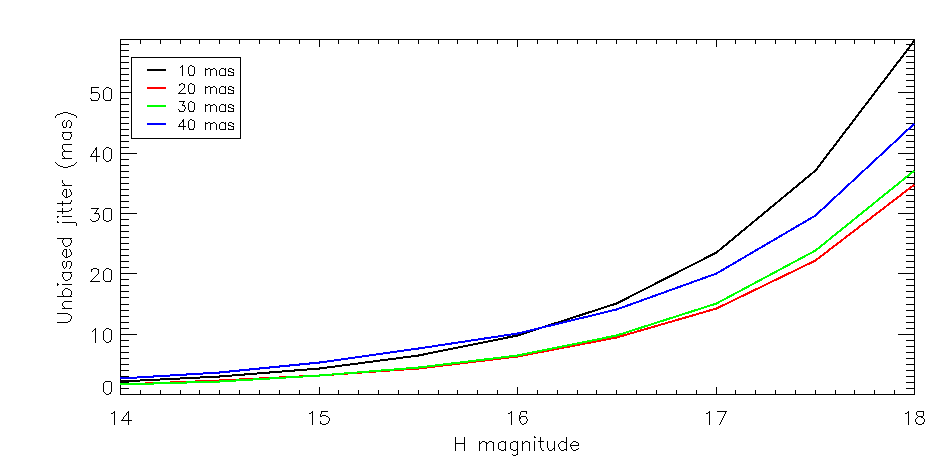}}
    \caption{Measurement noise tilt jitter in mas as a function of the H band magnitude.\label{fig:noise_LO}}
\end{figure}

We present here the study on the LO WFS pixel scale: we want to find the best pixel scale for both configurations of the LO WFS the 2$\times$2 and 1$\times$1 and we use an analytical approach to evaluate the measurement noise error for different flux levels and pixel scales.
We consider either a 1$\times$1 or 2$\times$2 configuration in two cases:
\begin{itemize}
    \item NGS close to the axis (best): SR(H) = 0.8 and Frame rate = 500 Hz.
    \item NGS at the edge of the technical FoV (worse): SR(H) = 0.2 and Frame rate = 100 Hz.
\end{itemize}
We report the noise on the centroid in Fig. \ref{fig:noise_LO}.
In both considered cases the pixel scale of 20 mas for the 1$\times$1 (40 mas for the 2$\times$2) provides the best result.
We also see that we can accept a slightly larger pixel scale without a significant loss of performance.
A tolerance of 10\% is considered.

\subsection{Vibrations}

\begin{figure}[htpb]
    \centering
    \subfigure[Cumulated PSDs of the first 10 modes for the three simulated cases.\label{fig:cum_PSD_vib}]
    {\includegraphics[width=0.42\columnwidth]{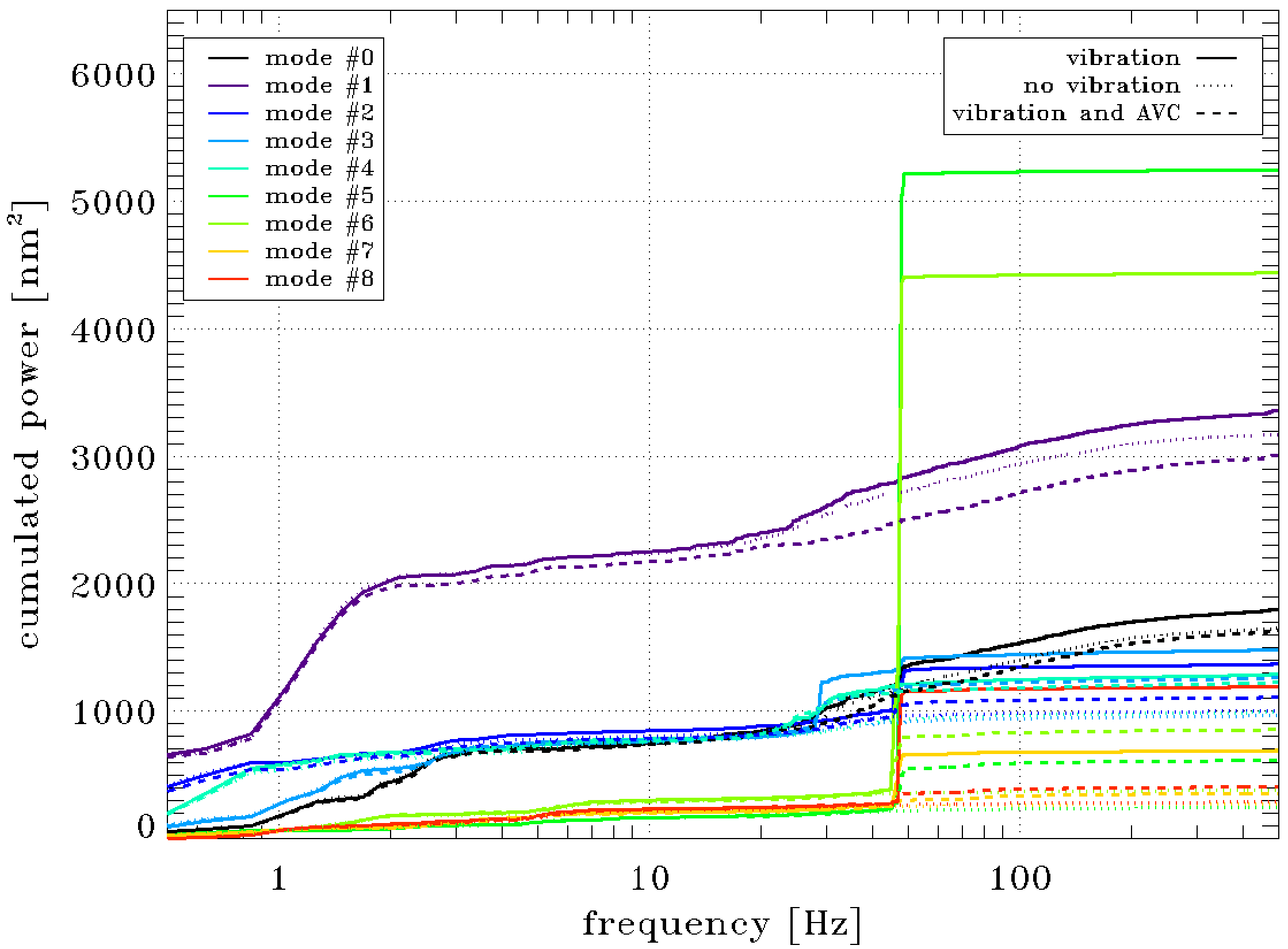}}
    \subfigure[On-axis RMS of turbulence and residual projected on a modal base.\label{fig:modal_plot_vib}]
    {\includegraphics[width=0.42\columnwidth]{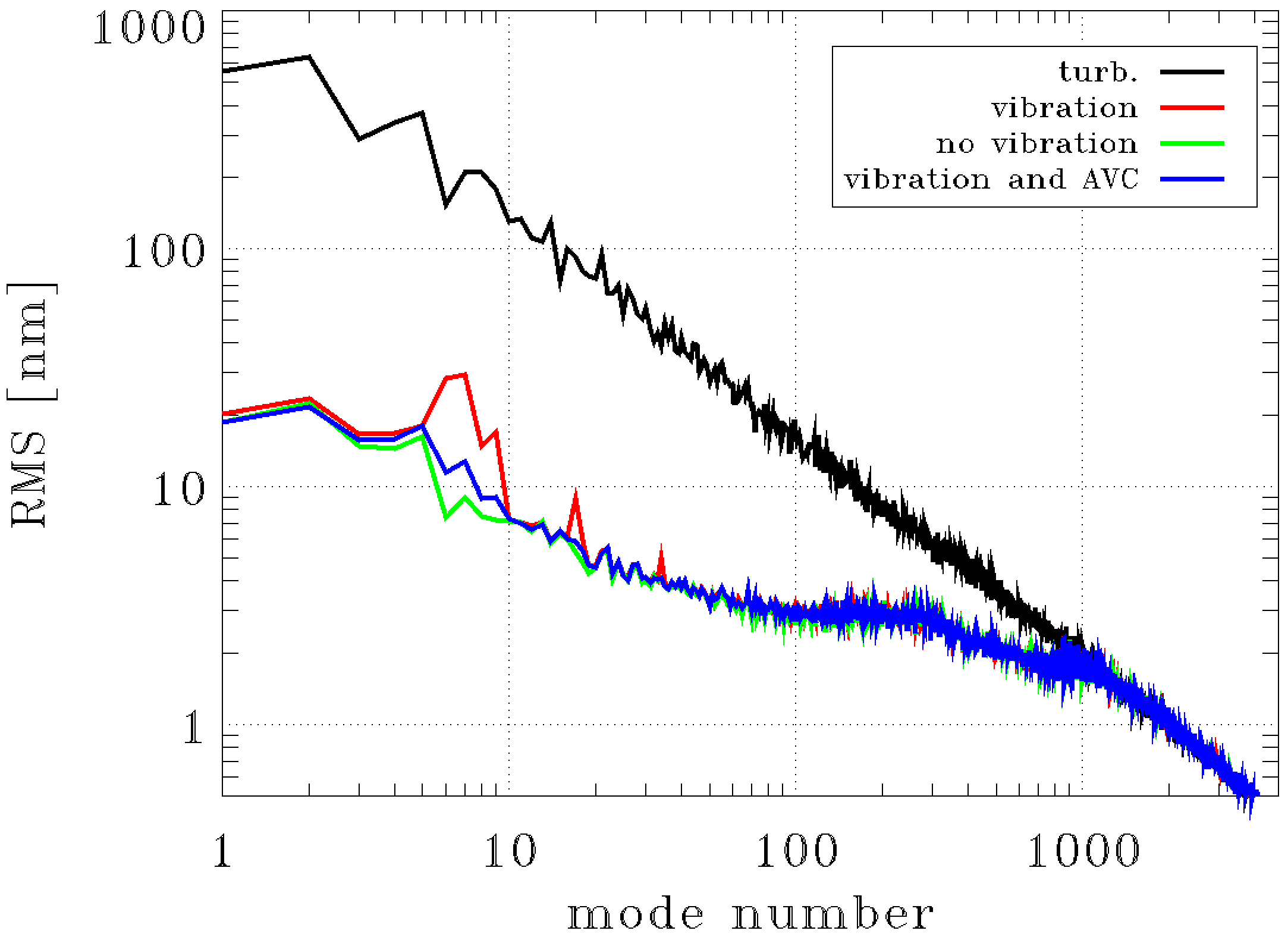}}
    \caption{Comparison between case without vibrations, with vibrations and with vibrations and the adaptive vibration cancellation algorithm. Modal base used here is made by 5 Zernikes and 4000 Karhunen-Lo{\'e}ve modes\cite{Wang:78}.\label{fig:vibrations}}
\end{figure}

We estimated the level of vibrations coming form ESO VLT from a set of closed loop PSDs of VLT-UT4 NEAR data (full telemetry data collected on 2016-09 at UT4 Cassegrain by the NEAR WFS working at 1kHz).
Modal PSDs show strong peaks at $\sim$48Hz in particular on trefoils (see solid lines in Fig. \ref{fig:cum_PSD_vib}). 
We used this data set to model the vibrations in the MAVIS simulations introducing more than 20 peaks.
Then we used the adaptive vibration cancellation (AVC) algorithm\cite{7320616} developed at ESO to compensate for them.
The simulation with vibrations showed an additional error of 50nm that is partially recovered by the AVC: this method is capable of reducing this error from 50 to 21nm (see Fig. \ref{fig:modal_plot_vib})).

\subsection{Mis-registrations}

\begin{figure}[htbp]
    \centering
    \includegraphics[width=0.5\linewidth]{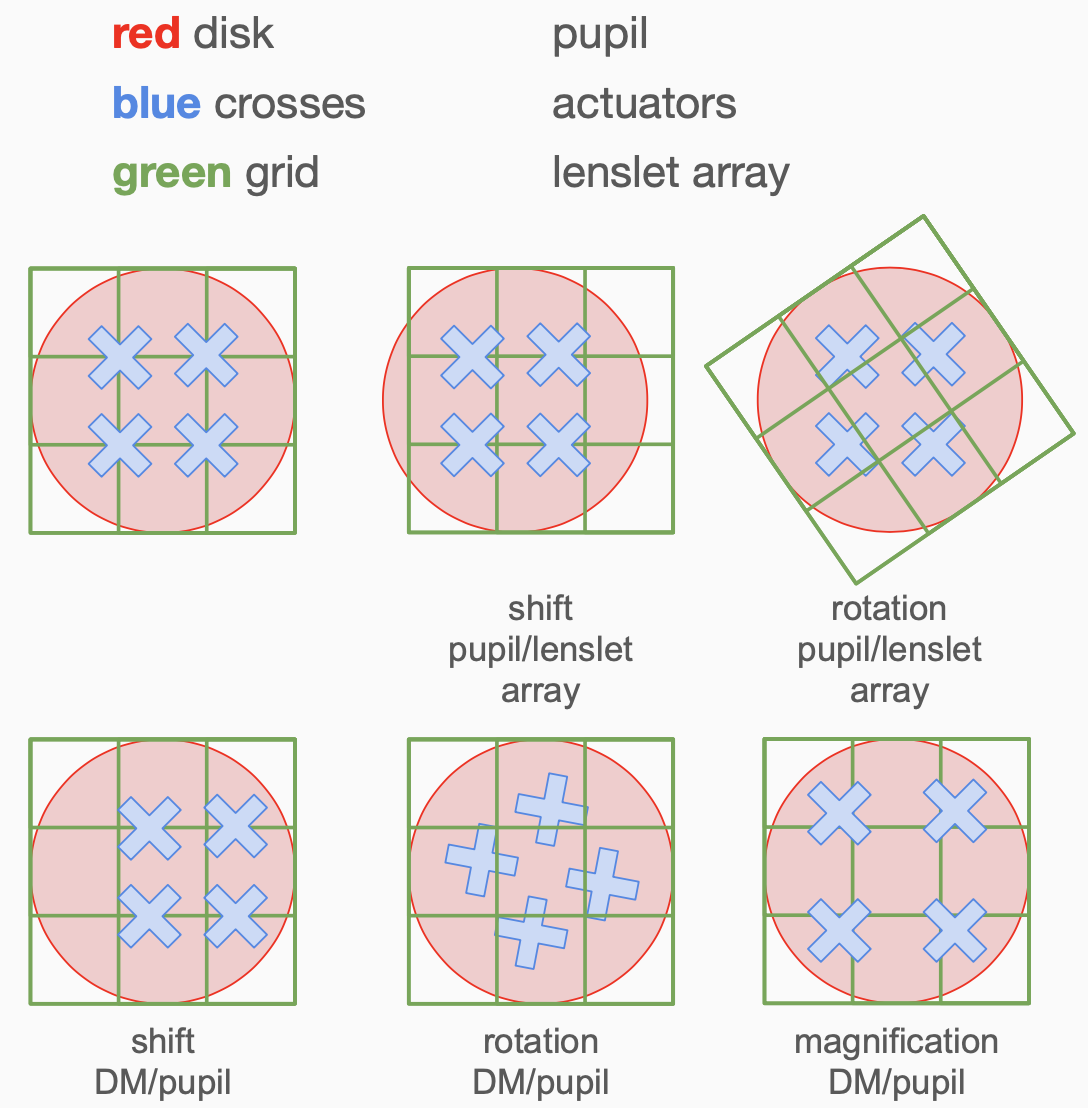}
    \caption{Legend of possible mis-registrations.}
    \label{fig:misal_legend}
\end{figure}

\begin{figure}[htbp]
    \centering
    \includegraphics[width=0.66\linewidth]{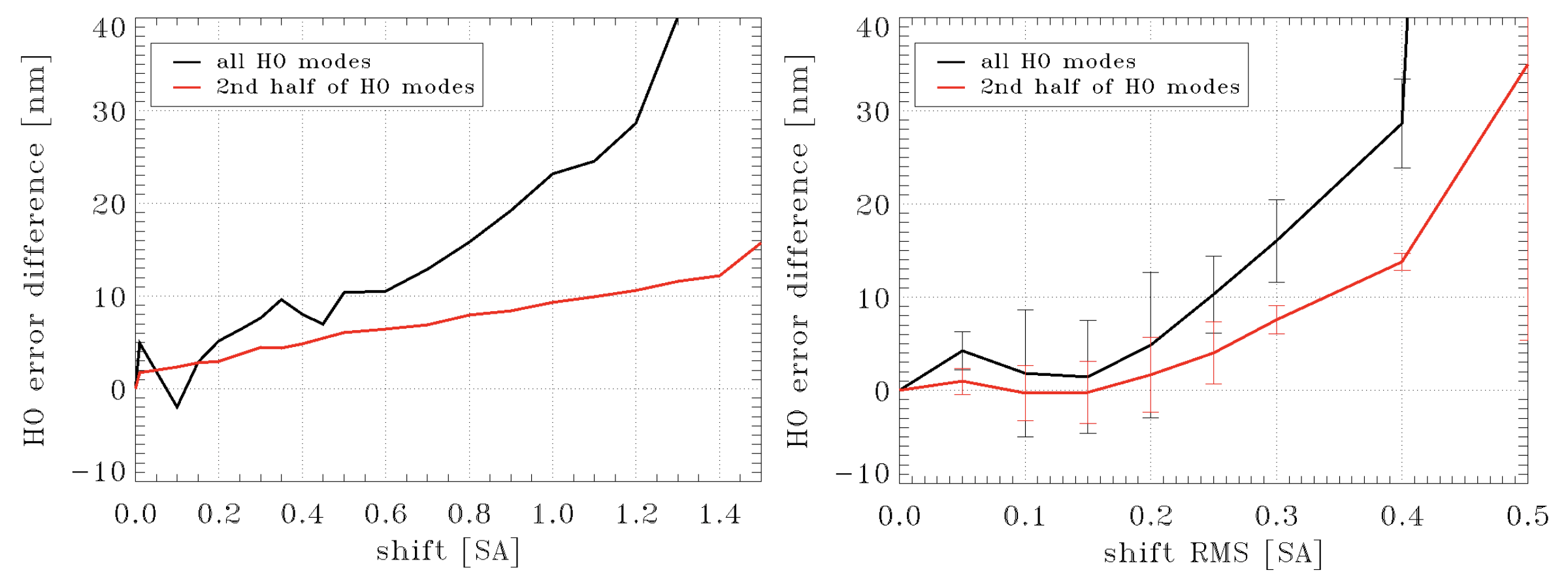}
    \caption{Sensitivity to shift pupil/lenslet array in sub-apertures (SA). Left, shift of a single WFS (note that for a value of 1.4SA the system is unstable), right, shift of all WFSs.}
    \label{fig:shift}
\end{figure}

\begin{figure}[htbp]
    \centering
    \includegraphics[width=0.66\linewidth]{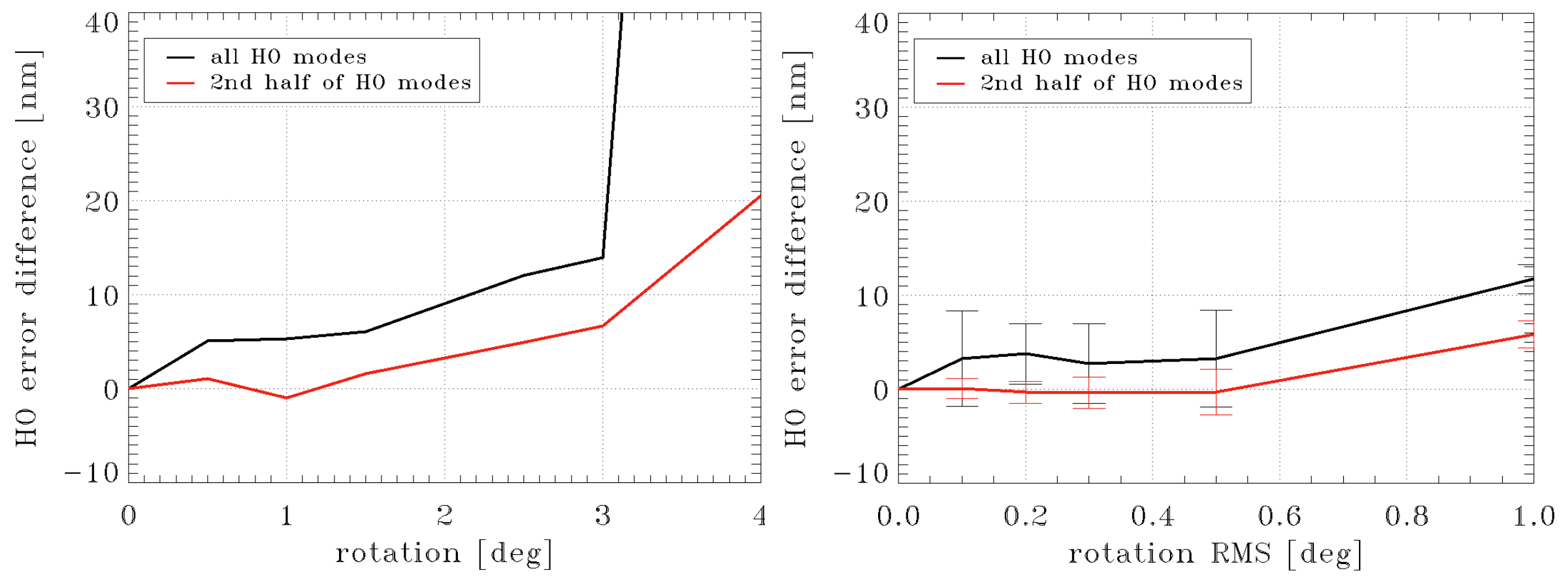}
    \caption{Sensitivity to rotation pupil/lenslet array. Left, rotation of a single WFS (note that for a value of 4deg the system is unstable), right, rotation of all WFSs.}
    \label{fig:rotation}
\end{figure}

\begin{figure}[htbp]
    \centering
    \includegraphics[width=0.99\linewidth]{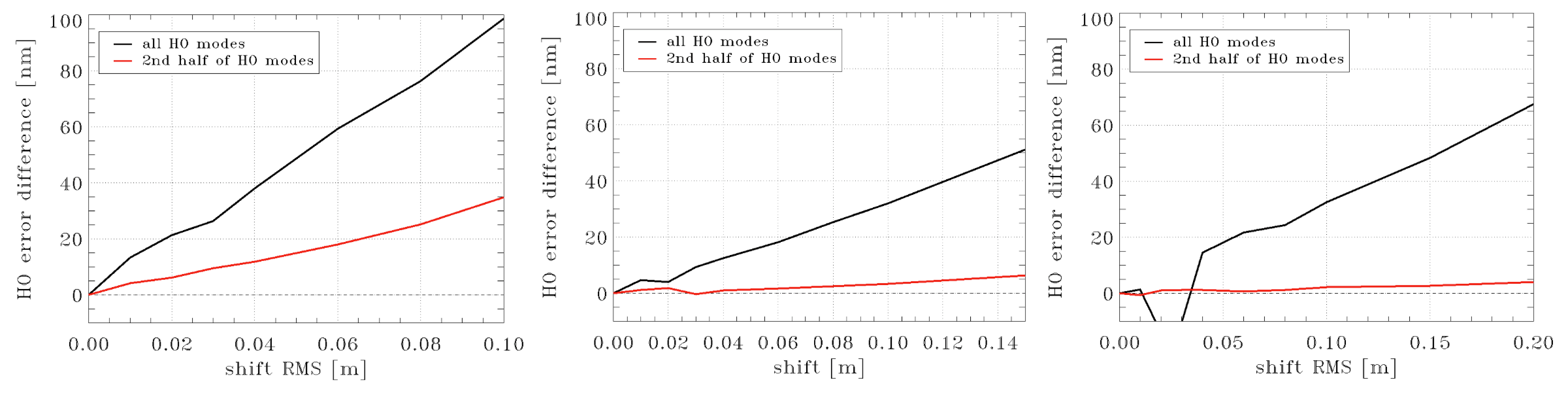}
    \caption{Sensitivity to shift DM/pupil for, left, DSM (note that this is not realistic because DSM is also the pupil), center, medium altitude post focal DM, right high altitude post focal DM.}
    \label{fig:DM_shift}
\end{figure}

\begin{figure}[htbp]
    \centering
    \includegraphics[width=0.99\linewidth]{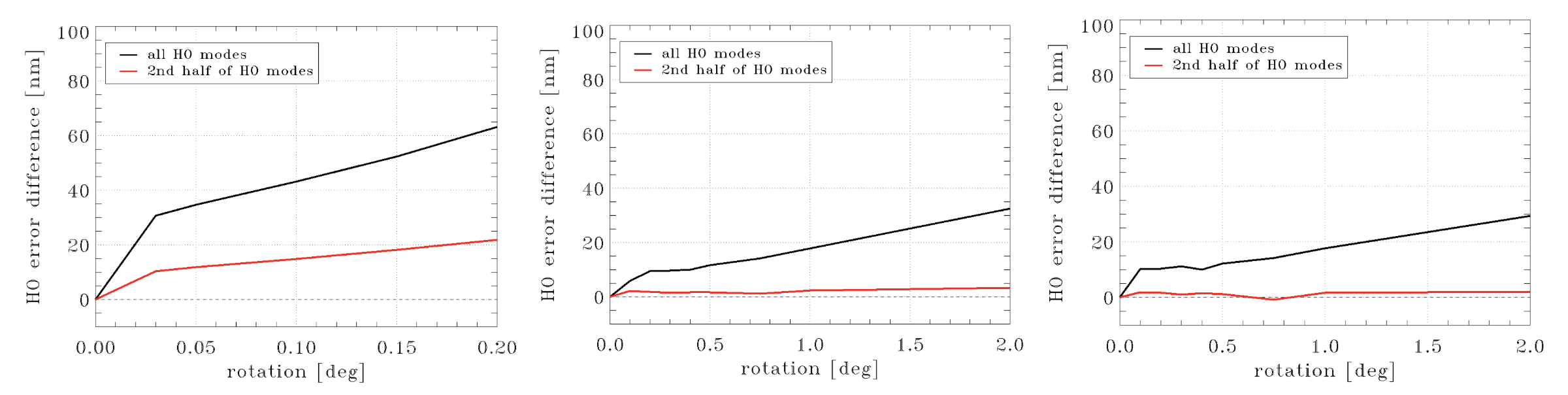}
    \caption{Sensitivity to rotation DM/pupil for, left, DSM, center, medium altitude post focal DM, right high altitude post focal DM.}
    \label{fig:DM_rotation}
\end{figure}

\begin{figure}[htbp]
    \centering
    \includegraphics[width=0.99\linewidth]{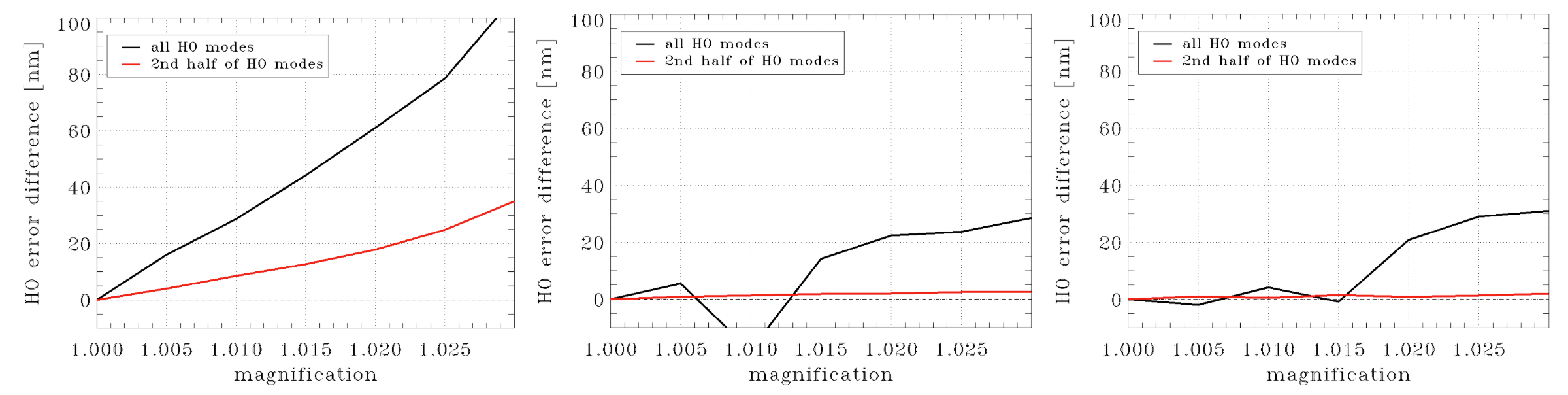}
    \caption{Sensitivity to magnification DM/pupil (the values are the ratio between DM and meta-pupil size) for, left, DSM (note that this is not realistic because DSM is also the pupil), center, medium altitude post focal DM, right high altitude post focal DM.}
    \label{fig:DM_magnification}
\end{figure}

Here we report sensitivity analysis to registration/alignement errors.
We consider that these mis-registrations are not calibrated so the reconstruction and projection matrices are the nominal ones.
We analysed 5 kinds of mis-registration errors that are shown in Fig. \ref{fig:misal_legend}.
The results are summarized in Fig. \ref{fig:shift}, \ref{fig:rotation}, \ref{fig:DM_shift}, \ref{fig:DM_rotation} and \ref{fig:DM_magnification}.
Please note that in the figures red line shows the sensitivity of the second half of DM modes and it is reported to focus on the effect of mis-registration on the higher spatial frequencies.
These results are used by the system engineering to set up requirements on the alignment (static and dynamic parts) pupil/DMs/WFSs.


\section{Conclusion}

MAVIS design is progressing and now we have a more accurate performance estimation than a year and a half ago: we refined our tools for system modelling and performance prediction, we reviewed and updated the control strategy and a few parameters, like total delay and laser spots geometry, and, finally, we explored the performance sensitivity to many parameters.
In particular, now we are more confident that the system can meet the requirements and we have an as complete as possible view on MAVIS performance over a large set of conditions.
In this phase performance estimation is not used only to verify top level requirements but also to derive sub-system requirements: this is the case of the study on the effect of mis-registrations.

\bibliography{biblio}
\bibliographystyle{spiebib}

\end{document}